\documentclass[aps,prl,twocolumn,superscriptaddress,showpacs,showkeys]{revtex4}

\usepackage{graphicx}

\begin{document}

\title{Commensurate spin dynamics in the superconducting state \\ 
of an electron-doped cuprate superconductor }

\author{K. Yamada}
\email[]{yamada@scl.kyoto-u.ac.jp}
\affiliation{Institute for Chemical Research, Kyoto University, Uji 611-0011, Japan}

\author{K. Kurahashi}
\altaffiliation{Kohzu Precision Co. Ltd., Setagaya, Tokyo 154-0005, Japan}
\affiliation{Department of Physics, Tohoku University, Aramaki Aoba, Sendai 980-8577, Japan}

\author{T. Uefuji}
\affiliation{Institute for Chemical Research, Kyoto University, Uji 611-0011, Japan}

\author{M. Fujita}
\affiliation{Institute for Chemical Research, Kyoto University, Uji 611-0011, Japan}

\author{S. Park}
\affiliation{Center for Neutron Research, National Institute of Standards and Technology, 
Gaithersburg, Maryland 20899}
\affiliation{Department of Materials and Nuclear Engineering,University of Maryland, College Park, Maryland 20742}

\author{S.-H. Lee}
\affiliation{Center for Neutron Research, National Institute of Standards and Technology, 
Gaithersburg, Maryland 20899}

\author{Y. Endoh}
\affiliation{Institute of Materials Research, Tohoku University, Sendai 980-8577, Japan}

\date{\today}

\begin{abstract}
We report neutron scattering studies on single crystals of the electron-doped ({\it n}-type) superconducting cuprate Nd$_{2-{\it x}}$Ce$_{\it x}$CuO$_{4}$ ({\it x}=0.15) with {\it T}$_{c}$ = 18 K and 25 K. Unlike the hole-doped ({\it p}-type) superconducting cuprates where {\it incommensurate} magnetic fluctuations commonly exist, the {\it n}-type cuprate shows {\it commensurate} magnetic fluctuations at the tetragonal (1/2 1/2 0) reciprocal points both in the superconducting and in the normal state. A spin gap opens up when the {\it n}-type cuprate becomes superconducting, as in the optimally doped {\it p}-type La$_{2-{\it x}}$Sr$_{\it x}$CuO$_{4}$. The gap energy, however, increases gradually up to about 4 meV as {\it T} decreases from {\it T}$_{c}$ to 2 K, which contrasts with the spin pseudogap behavior with a {\it T}-independent gap energy in the superconducting state of {\it p}-type cuprates. 
\end{abstract}

\pacs{74.25.Ha, 74.72.-h, 74.72.Jt}

\keywords{High-$T_{{\rm c}}$ superconductivity, Electron-doped system, Neutron scattering}

\maketitle

High-{\it T}$_{c}$ superconductivity emerges when charge carriers, holes or electrons, are doped into an antiferromagnetic Mott insulator \cite{Bednorz,Kastner,Tokura}. Mechanism of the superconductivity lies on their common two-dimensional CuO$_{2}$ planes into which the charge carriers go. One of the issues in understanding the mechanism is question of the electron-hole symmetry. Electronic structure of the optimally doped cuprates shows evidence for the electron-hole symmetry. \cite{Sato, Armitage} Their phase diagrams over doping concentrations, however, are asymmetric. \cite{Damascelli} For hole doping, antiferromagnetism rapidly weakens and is replaced by a spin-glass-like phase with characteristics of incommensurate spin correlations and pseudo-gap in transport measurements. The pseudo-gap temperature, {\it T}$^{\ast}$, is well defined in the underdoped region and decreases with doping. The system becomes superconducting (SC) over a wide range of the hole concentration, {\it x}, around the optimal {\it x} = 0.15. The SC state has incommensurate spin correlations with a {\it T}-independent spin gap. The normal state of the underdoped and optimally doped SC region also shows unusual non-Fermi-liquid (FL) behaviors. There are increasing evidence for a Quantum Critical Point (QCP) around the optimal doping which is responsible for the unusual properties of the SC and the normal phase \cite{Tallon,Tallon2,Kusko,Panagopoulos}. For the electron-doped ({\it n}-type) cuprates, on the other hand, antiferromagnetism survives until the superconductivity appears over a narrow range of {\it x} around the optimal {\it x} $\sim$ 0.15. The normal state of the {\it n}-type cuprates shows Fermi-Liquid {\it T}$^{2}$ behavior in resistivity rather than the linear behavior of the hole-doped ({\it p}-type) cuprates. Therefore, investigating similarities and differences of the {\it n}-type and {\it p}-type cuprates would be crucial to understanding physics of the high-{\it T}$_{c}$ superconductivity. Compared to a large number of studies on the hole-doped cuprates using various techniques, however, only a small number of key experiments have been done on electron-doped cuprates \cite{Sato,Armitage,Tsuei,Yanase} mainly because it is difficult to grow large single crystals and to prepare homogeneous superconducting samples by post-growth heat treatment. 

In this paper, we report neutron scattering measurements on single crystals of the electron-doped ({\it n}-type) superconducting cuprate Nd$_{2-{\it x}}$Ce$_{\it x}$CuO$_{4}$ ({\it x} = 0.15) with {\it T}$_{c}$ = 18 K and 25 K. We have found {\it commensurate} magnetic fluctuations at the tetragonal (1/2 1/2 0) reciprocal points in the superconducting and normal states. A spin gap opens up when the system becomes superconducting, as in the optimally doped {\it p}-type La$_{2-{\it x}}$Sr$_{\it x}$CuO$_{4}$ \cite{Yamada,CHLee,Lake1}. The gap energy, however, increases gradually up to about 4 meV as {\it T} decreases from {\it T}$_{c}$ to 2 K, which contrasts with the {\it T}-independent spin pseudogap energy for the superconducting state of the {\it p}-type cuprate near {\it T}$_{c}$. Our results indicate that doped electrons self-organize in a different way than holes that form stripes such as in (La,Nd)$_{2-{\it x}}$Sr$_{\it x}$CuO$_{4}$ \cite{Tranquada,Zhou}, La$_{2-{\it x}}$(Sr,Ba)$_{\it x}$CuO$_{4}$ \cite{Fujita} and the isostructural insulator La$_{2-{\it x}}$Sr$_{\it x}$NiO$_{4}$ \cite{Tranquada2,SHLee}. 

Sizable single crystals of NCCO with {\it x} = 0.15 were grown by a traveling-solvent-floating-zone (TSFZ) method. The as-grown crystal is an antiferromagnetic insulator with the N{\' e}el temperature {\it T}$_{N}$ of around 125 $\sim$ 160 K depending on the excessive oxygen concentration in the crystal. Bulk superconductivity only appears with proper heat treatments on these as-grown crystals. Detailed procedure of the crystal growth and the heat treatment is described in a separate paper \cite{Kurahashi}. For the SC sample, {\it T}$_{c}$ is determined from zero-field-cooled diamagnetic susceptibilities measured by a SQUID magnetometer. For the present study, we used two SC samples with {\it T}$_{c}$ = 18 K and 25 K.

Neutron scattering experiments were performed on the thermal neutron triple-axis spectrometer of Tohoku University, TOPAN and a cold neutron triple-axis spectrometers of University of Tokyo, HER installed at JRR-3M in JAERI, Tokai Establishment. We performed the experiment with low energy excitations below 1 meV on the cold neutron triple-axis spectrometer, SPINS at the National Institute of Standards and Technology (NIST) Center for Neutron Research.  Incident neutron energies of 13.7 meV for TOPAN and 5 meV for HER and SPINS were selected using the (0 0 2) reflection of pyrolytic graphite monochromators. Additionally, in order to eliminate the higher-order reflected beams, pyrolytic graphite filter for thermal neutrons and Be filter for cold neutrons were placed in the up or down stream of the sample position. Previously, a few attempts have been made to study spin dynamics in NCCO using neutron scattering technique \cite{Matsuda}. However, no well-defined magnetic signal has been found in the superconducting phase. In the previous measurements, the total volume of the sample was 0.5 cc and the ({\it hhl}) scattering plane was examined. For our study, we have investigated mostly the ({\it hk}0) scattering plane to increase the signal from two dimensional magnetic rod along c-axis by using the larger vertical angular divergence of the beam. Furthermore, we cut a long single crystalline rod with a total volume of 2 cc into three pieces and stacked those vertically using an aluminum sample-holder. The holder was mounted in an aluminum can attached either to the cold plate of a $^{4}$He-closed cycle refrigerator, or to a top-loading liquid-He cryostat, which cools down to 1.5 K. 

Figure 1 (a) shows temperature dependence of elastic neutron scattering intensity at (3/2 1/2 0) reflection, obtained from three samples: the as-grown insulating sample and the two heat-treated superconducting samples with {\it T}$_{c}$ = 18 K and 25 K. For the as-grown insulating sample, upon cooling the intensity starts gradually increasing below {\it T}$_{N}$ $\sim$ 140 K, signaling the antiferromagnetic long range order of Cu$^{2+}$ moments. And it rapidly increases below {\it T}$_{Nd}$ $\sim$ 20 K due to the participation of Nd$^{3+}$ moments. When the sample gets the proper heat-treatment and becomes superconducting, the static magnetic order is drastically suppressed. Even though the magnetic elastic peaks remain even in the SC phase, the superconducting samples have lower {\it T}$_{N}$ $\sim$ 60 - 80 K and {\it T}$_{Nd}$, and weaker intensity than the insulating sample. The suppression is more severe for the superconducting sample with the higher {\it T}$_{c}$ = 25 K than for the one with {\it T}$_{c}$ = 18 K. The width of the static peak was resolution limited for the insulating sample and was broadened for the superconducting samples. By fitting the static peak with a Lorentzian convoluted with the instrumental resolution, we obtained the in-plane and the out-of-plane correlation length, $\xi_{ab}$ = 150(60) {\rm \AA} and $\xi_{c}$ = 80(30) {\rm \AA} respectively, at {\it T} = 8 K for the {\it T}$_{c}$ = 25 K sample. This indicates that superconductivity competes with the magnetic order in the {\it n}-type cuprate as in {\it p}-type cuprates.

\begin{figure}
\includegraphics[scale=0.65]{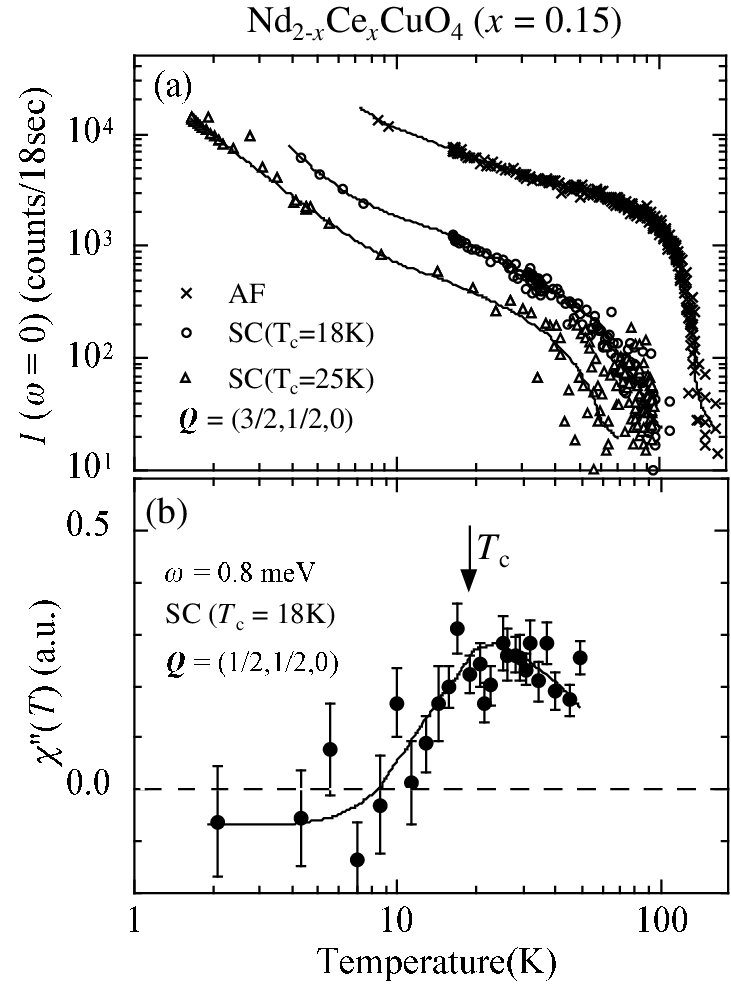}
\caption{\label{1} (a) {\it T}-dependence of the elastic magnetic peak intensities of NCCO ({\it x} = 0.15) at (3/2 1/2 0) measured from an as-grown insulating (cross) and the two reduced superconducting single crystals with {\it T}$_{c}$ = 18 K (open circles) and 25 K (open triangles). (b) {\it T}-dependence of dynamical susceptibility $\chi''(\omega$ = 0.8 meV) at (1/2,1/2,0) obtained from the reduced sample of NCCO ({\it x} = 0.15, {\it T}$_{c}$ =18 K).}
\end{figure}

To investigate the relationship between the superconductivity and the dynamic spin correlations, we have performed constant energy scans around the antiferromagnetic (1/2 1/2 0) point. Figures 2 (a) and (b) show the results for the normal state obtained with $\omega$ = 2 meV along the [1 0 0] and [1 1 0] directions, which are parallel and diagonal to Cu-O bonds in the CuO$_{2}$ plane, respectively. In both directions, a commensurate peak appears. This results sharply contrast with the incommensurate peaks found in the {\it p}-type superconducting cuprates\cite{Tranquada}. Solid lines are the best fits to a gaussian convoluted with the instrumental resolution function yielding the intrinsic Half-Width-of-Half-Maximum (HWHM) of 0.025(3) a$^{\ast}$. Indeed, the data shown in Figures 2 (a) and (b) are the first direct experimental evidence for the commensurate spin fluctuations in the {\it n}-type cuprate. The peak-widths of the SC samples are substantially broader than those of the AF phase, while the {\it q} - integrated peak intensities are comparable between the AF and SC samples except at low temperatures below {\it T}$_{c}$. Furthermore, the peak-width is broader for the {\it T}$_{c}$ = 25 K sample than the {\it T}$_{c}$ = 18 K sample. Figure 1 (b) shows that for $\omega$ = 0.8 meV the commensurate spin fluctuations diminishes as {\it T} decreases below {\it T}$_{c}$, which indicates opening of a spin gap in the superconducting state. Figures 2 (c) and (d) clearly show the depletion of the spectral weight for $\omega <$ 3 meV. To study the energy dependence of the dynamic spin fluctuations in detail, we have performed the constant $\omega$-scans shown in Figure 2 with various energy transfers. The data were fitted to a gaussian convoluted with the instrumental resolution function. The integrated intensity of the gaussian was converted to the imaginary part of dynamic susceptibility via the fluctuation dissipation theorem, $I(\omega) \propto \chi''(\omega) [1+n(\omega)]/\pi$ where $n(\omega)$ is the Bose thermal population factor. Figure 3 shows the resulting $\chi''(\omega)$ as a function of the energy transfer, $\omega$. For both SC samples, at {\it T} = 2 K, $\chi''(\omega)$ has a gap of 2$\Delta \sim$ 4 meV and 5 meV for the {\it T}$_{c}$ = 18 K and 25 K samples, respectively. The fact that the sample with the higher {\it T}$_{c}$ has the larger 2$\Delta$, whereas it has the weaker Nd-ordering and the lower {\it T}$_{Nd}$, tells us that the gapped commensurate spin fluctuation is an intrinsic property of the spin dynamics in the optimally doped {\it n}-type superconductor. Such a spin gap has also been reported in the {\it p}-type La$_{2-{\it x}}$Sr$_{\it x}$CuO$_{4}$ near optimal doping. A spin gap of 2$\Delta$ = 6 $\sim$ 7 meV was observed in the optimally-doped La$_{2-{\it x}}$Sr$_{\it x}$CuO$_{4}$ \cite{CHLee,Lake1}. As shown in the inset of Figure 4, their maximum gaps 2$\Delta_{max}$ behave linearly with the SC temperature scale $Ck_{B}T_{c}$ with $C = 1.9$, irrespective of carrier type. There is, however, a qualitative difference of the spin dynamics between the {\it n}-type and the {\it p}-type cuprates: For the {\it n}-type Nd$_{2-{\it x}}$Ce$_{\it x}$CuO$_{4}$ ({\it x} = 0.15), as shown in Figure 3, upon warming $\chi''({\bf Q},\omega)$ shifts toward lower energies and eventually fills up the gap in the normal phase. The shift of the spectral weight in the energy spectrum is summarized in Figure 4: 2$\Delta$ decreases with increasing {\it T} and eventually disappears when the system enter the normal phase. This suggests the absense or degradation of spin pseudogap state and is in a sharp contrast with those observed in the La$_{2-{\it x}}$Sr$_{\it x}$CuO$_{4}$ ({\it x} = 0.15) in which for {\it T} $\ll$ {\it T}$_{c}$ $\chi''({\bf Q},\omega)$ diminishes for all $\omega < 2\Delta$, and as {\it T} approaches {\it T}$_{c}$, $\chi''({\bf Q},\omega)$ smears into lower $\omega$ without remarkable change in 2$\Delta$ \cite{Yamada,CHLee}.

\begin{figure}
\includegraphics[scale=0.40]{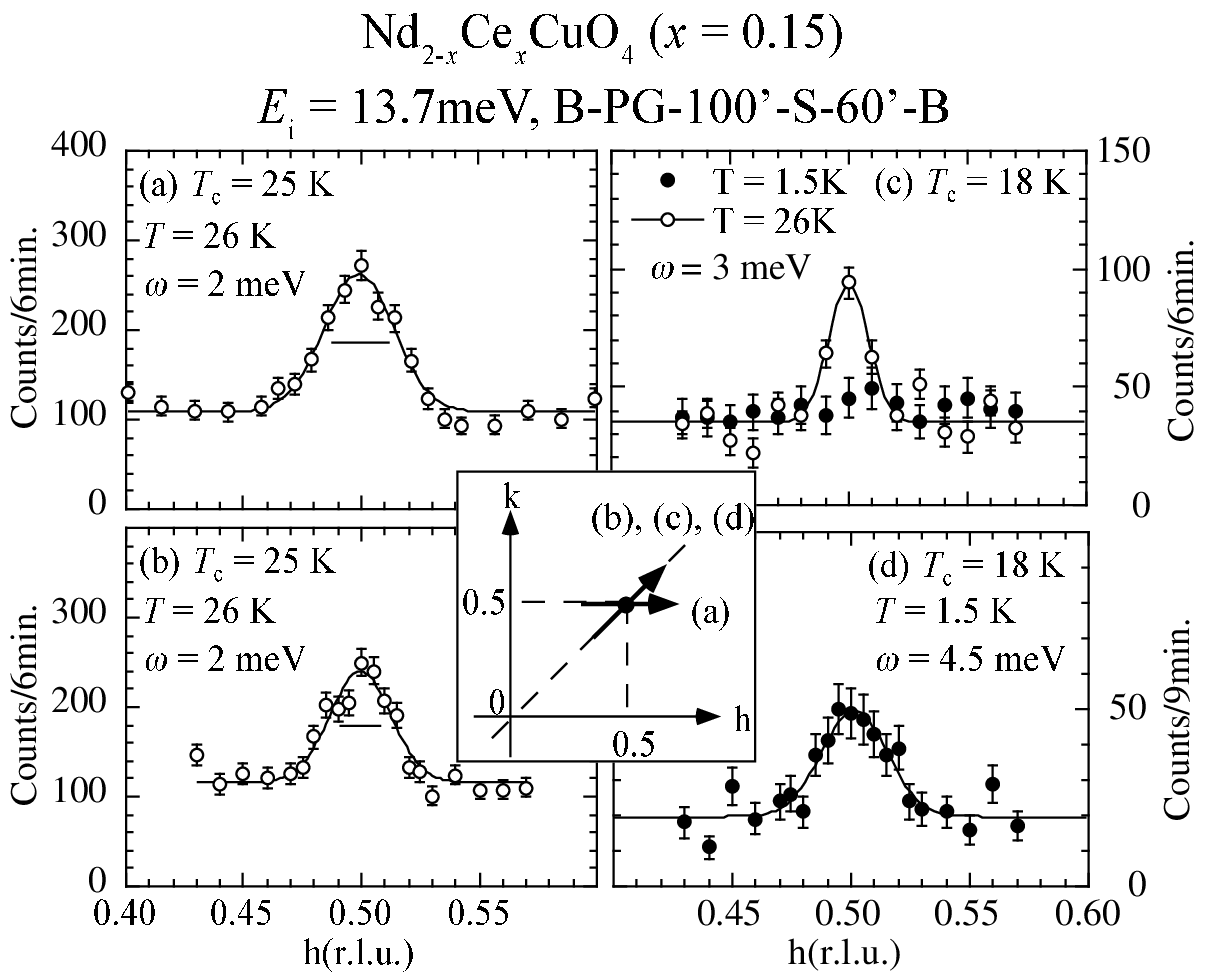}
\caption{\label{2} Constant-$\omega$ scans around (1/2,1/2,0) obtained from the reduced sample of NCCO ({\it x} = 0.15, {\it T}$_{c}$ =18 K) in the normal ({\it T} = 26K) and superconducting ({\it T} = 1.5 K) phases. (a) 26 K, normal, $\omega$ = 2 meV along (0,{\it k},0) direction. (b) 26 K, normal, $\omega$ = 2 meV along ({\it h},{\it h},0) direction. (c) 26 K, normal (open circles) and 1.5 K, superconducting (filled circles), with $\omega$ = 3 meV along (0,{\it k},0) direction. (d) 26 K, normal, $\omega$ = 2 meV along (0,{\it k},0) direction. Horizontal bars indicate the Full-Width-of-Half-Maximum of the instrumental resolution. Solid lines are explained in the text.}
\end{figure}

\begin{figure}
\includegraphics[scale=0.65]{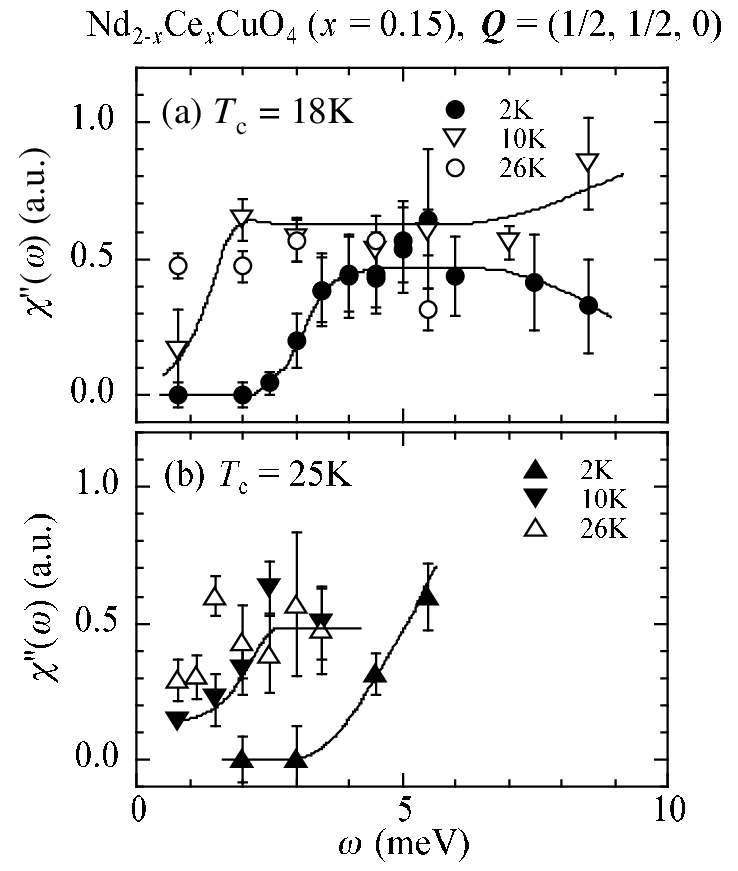}
\caption{\label{3} Energy spectra of $\chi''(\omega)$ obtained from the normal and superconducting phases (a) of NCCO ({\it x} = 0.15, {\it T}$_{c}$ =18 K) and (b) of NCCO ({\it x} = 0.15, {\it T}$_{c}$ = 25 K).}
\end{figure}

Commensurate short range spin correlations in the superconducting phase of the {\it n}-type cuprate suggest that the doped electrons may be inhomogeneously distributed or form droplets/bubbles in the CuO$_{2}$ planes, rather than organizing into one-dimensional stripes as the doped holes seems to do in the {\it p}-type cuprates. Their different orbital characters might be responsible for the different behaviors. Doped holes are dominantly introduced into the 2$p_{xy}$ orbital of oxygen ions in CuO$_{2}$ planes and induce the frustrated magnetic interactions between the neighboring Cu spins which stabilize the formation of stripes. The elastic incommensurate magnetic signal increases when an external magnetic field is applied perpendicular to the CuO$_2$ planes.\cite{Katano,Lake2,Khaykovich} On the other hand, doped electrons predominantly enter into the 3$d_{x^{2}-y^{2}}$ orbitals of Cu ions to make the Cu site nonmagnetic, which reduces the size of the antiferromagnetic domains without changing the commensurability. For the {\it p}-type SC cuprates the stripe phase (or the pseudo-gap phase) competes with the superconducting phase, whereas for the {\it n}-type SC cuprates it is the three-dimensional antiferromagnetic state that competes with the superconductivity. Our results show that for {\it n}-type cuprates the AFM coexists with superconductivity at the optimal region. This indicates that the transition from the AFM to superconductivity upon doping is a first order in nature and therefore it lacks a QCP. This difference may be responsible for their different properties in SC and normal states\cite{Sachdev,Varma,Coleman}. 
\begin{figure}
\includegraphics[scale=0.50]{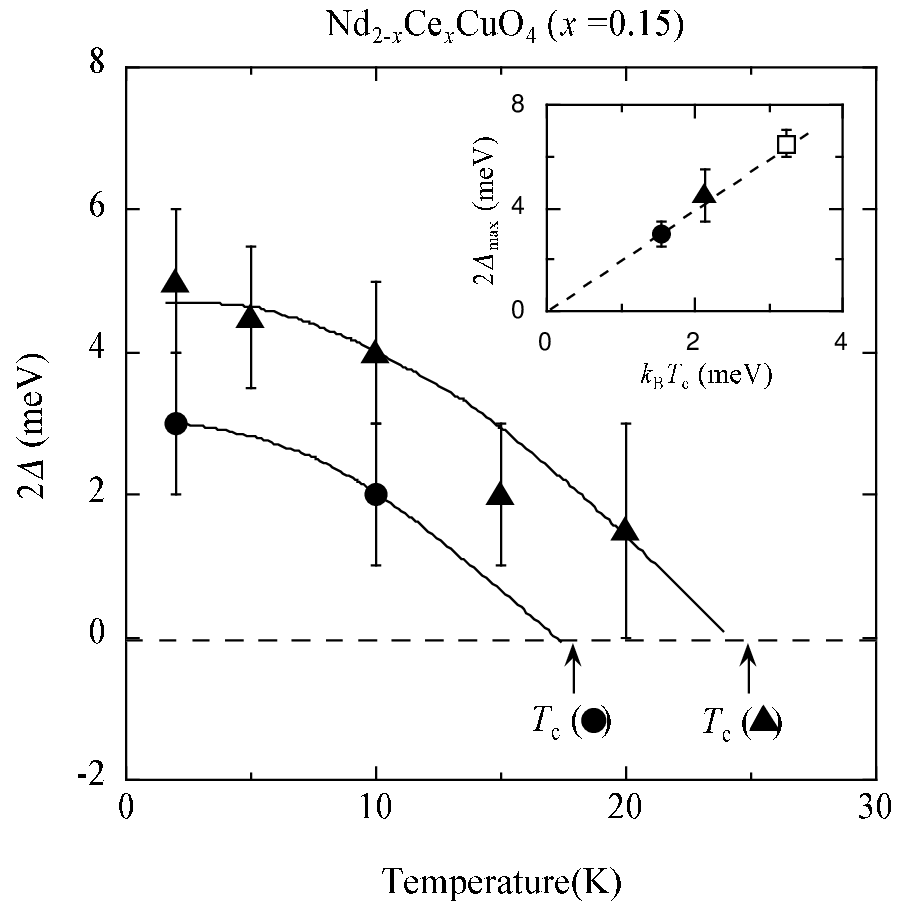}
\caption{\label{4} {\it T}-dependence of the spin gap, 2$\Delta$ for the {\it n}-type superconducting cuprate NCCO ({\it x} = 0.15) with {\it T}$_{c}$ =18 K and 25 K. Lines are guide to eyes. The inset shows 2$\Delta$ versus $k_{B}T_{c}$ obtained from various high-{\it T}$_{c}$ cuprates. LSCO data were taken from \cite{CHLee}. At 26 K, the spin gaps are absent for both NCCO samples.}
\end{figure}

In summary, we revealed by neutron scattering study on the single crystals of Nd$_{1.85}$Ce$_{0.15}$CuO$_{4}$ the coexistence of dynamical gapped spin fluctuations with the electron-doped superconductivity.  In contrast to the incommensurate spin fluctuations in the hole-doped systems the spin fluctuations are commensurate to the CuO$_{2}$ square lattice. The energy gap of around 4 meV closes at or near {\it T}$_{c}$, which suggests the absence or degradation of the spin pseudo gap state in the electron-doped superconductor. More comprehensive study on the electron-doped superconductor free from the effect of rare-earth magnetic moments is highly required for the fully understanding of the universal properties of spin fluctuations irrespective of types of carrier.

\begin{acknowledgments}
We thank Y. Kojima, I. Tanaka, S. Hosoya, K. Hirota, H. Y. Kee, J. M. Tranquada, P. A. Lee, G. Shirane and R. J. Birgeneau for their valuable discussions. This work was supported by the Japanese Ministry of Education, Culture, Sports, Science and Technology, Grant-in-Aid for Scientific Research on Priority Areas Contract, Scientific Research (A), Encouragement of Young Scientists, Creative Scientific Research, the Japan Science and Technology Corporation, the Core Research for Evolutional Science and Technology Project (CREST), and the National Science Foundation under Agreement No. DMR-9986442.
\end{acknowledgments}

\end{document}